\documentclass{article}

\usepackage[dvipsnames]{xcolor}
\definecolor{mygray}{gray}{0.6}

\usepackage{hyperref}
\hypersetup{colorlinks=true,citecolor=mygray,linkcolor=mygray,urlcolor=cyan}

\usepackage{arxiv}
\usepackage[utf8]{inputenc}
\usepackage[T1]{fontenc}
\usepackage{url}
\usepackage{booktabs}
\usepackage{amsfonts}
\usepackage{nicefrac}
\usepackage{lipsum}
\usepackage{graphicx}
\usepackage[comma, sort&compress, numbers]{natbib}

\usepackage{setspace}

\title{Analyzing the Brain's Dynamic Response to Targeted Stimulation using Generative Modeling}

\author{ 
        {Rishikesan Maran\textsuperscript{1}, Eli J.~M\"uller\textsuperscript{1} and Ben D.~Fulcher\textsuperscript{1}} \\
        \\
        \textsuperscript{1}School of Physics, University of Sydney, Camperdown 2006, NSW, Australia.
}

\renewcommand{\shorttitle}{Modeling the Brain's Response to Targeted Stimulation}

\hypersetup{
pdftitle={Analyzing the Brain's Dynamic Response to Targeted Stimulation using Generative Modeling},
pdfsubject={q-bio.NC, physics.bio-ph},
pdfauthor={Rishikesan Maran, Eli J.~M\"uller Ben D.~Fulcher},
pdfkeywords={brain stimulation, brain modeling, systems biology, complex systems},
}

\begin{document}
\maketitle

\begin{abstract}
Generative models of brain activity have been instrumental in testing hypothesized mechanisms underlying brain dynamics against experimental datasets.
Beyond capturing the key mechanisms underlying spontaneous brain dynamics, these models hold an exciting potential for understanding the mechanisms underlying the dynamics evoked by targeted brain-stimulation techniques.
This paper delves into this emerging application, using concepts from dynamical systems theory to argue that the stimulus-evoked dynamics in such experiments may be shaped by new types of mechanisms distinct from those that dominate spontaneous dynamics.
We review and discuss: (i) the targeted experimental techniques across spatial scales that can both perturb the brain to novel states and resolve its relaxation trajectory back to spontaneous dynamics; and (ii) how we can understand these dynamics in terms of mechanisms using physiological, phenomenological, and data-driven models.
A tight integration of targeted stimulation experiments with generative quantitative modeling provides an important opportunity to uncover novel mechanisms of brain dynamics that are difficult to detect in spontaneous settings.
\end{abstract}


\section{Introduction}

Statistical analyses of experimental neuroimaging data have long demonstrated that the brain supports a rich repertoire of spatiotemporal dynamics \citep{lopesdasilvaModelsNeuronalPopulations1976, biswalFunctionalConnectivityMotor1995, honeyNetworkStructureCerebral2007, vidaurreBrainNetworkDynamics2017}.
To better understand the intrinsic mechanisms which give rise to the brain's dynamical properties, generative models of brain activity---also known as computational models, dynamic models, or mathematical models---can be used to test hypothesized mechanisms against properties of brain dynamics that they aim to explain or predict.
While the term `generative model' can have a range of interpretations across fields, here we use it as shorthand for `generative model of brain dynamics', to describe a model that can generate (through simulation) a set of time-series of brain activity that can be compared to data measured experimentally \citep{ramezanian-panahiGenerativeModelsBrain2022}.
Existing generative models allow us to simulate brain activity based on a wide variety of encoded mechanisms across spatial scales \citep{breakspearDynamicModelsLargescale2017, decoDynamicBrainSpiking2008, cabralFunctionalConnectivityDynamically2017, kimLinearDynamicsControl2020, acharyaBrainModelingControl2022, ramezanian-panahiGenerativeModelsBrain2022}, allowing the hypothesized mechanisms to be directly evaluated against the dynamics of activity accessed in experimental neural recordings.

A crucial advantage that generative models provide is the ability to rigorously test candidate mechanisms against datasets from different experimental settings.
Early generative models were used to test mechanisms underlying dynamics of brain activity recorded during cognitive or motor tasks, or in response to sensory stimuli \citep{jirsaTheoreticalModelPhase1994, jirsaFieldTheoryElectromagnetic1996, jirsaDerivationMacroscopicField1997, zipserRecurrentNetworkModel1991, brunelEffectsNeuromodulationCortical2001, jansenElectroencephalogramVisualEvoked1995, rennieUnifiedNeurophysicalModel2002, davidMechanismsEvokedInduced2006, moranNeuralMassModel2007, kerrPhysiologybasedModelingCortical2008}.
Rich dynamical properties were later found even in \textit{spontaneous} brain activity, i.e., activity not attributable to any externally applied task or stimulus \citep{foxSpontaneousFluctuationsBrain2007}. 
Many of these properties have hypothesized mechanistic explanations that have been tested through generative models \citep{robinsonModalAnalysisCorticothalamic2001, robinsonDynamicsLargescaleBrain2002, breakspearUnifyingExplanationPrimary2006, decoDynamicsRestingFluctuations2017, freyerBiophysicalMechanismsMultistability2011, abdelnourFunctionalBrainConnectivity2018, honeyPredictingHumanRestingstate2009, nozariMacroscopicRestingstateBrain2023}.
As datasets from different experimental settings reveal novel spatiotemporal properties of brain dynamics, generative models present new opportunities to refine and hypothesize new underlying mechanistic accounts that can accurately capture them.

A new opportunity of growing interest to generative modeling, is to test our understanding of the dynamics of brain activity evoked by targeted stimulation.
Unlike traditionally used sensory stimulation paradigms, targeted stimulation techniques directly target neurons or neural populations using concentrated external inputs of energy .
This approach bypasses the constraints imposed by the brain's natural sensory pathways, offering significantly greater spatial precision in target selection and flexibility in stimulation strength.
Commonly used targeted stimulation techniques include optogenetics \citep{deisserothOptogenetics10Years2015}, electrode stimulation \citep{kringelbachTranslationalPrinciplesDeep2007, flesherIntracorticalMicrostimulationHuman2016}, transcranial magnetic stimulation \citep{rogaschAssessingCorticalNetwork2012}, and chemogenetics \citep{rothDREADDsNeuroscientists2016}, all of which are applicable \textit{in vivo}, and allow for simultaneous measurement of evoked responses with compatible measurement techniques \citep{osheaDirectNeuralPerturbations2022, markicevicCorticalExcitationInhibition2020, sanzeniMechanismsUnderlyingReshuffling2023, bonnardRestingStateBrain2016, luboeinskiNonlinearResponseCharacteristics2020}.
This review focuses on the new opportunities that datasets of activity measured from targeted stimulation experiments present in validating and refining our generative models of brain activity, and thereby better understanding the important mechanisms which underpin brain dynamics.

Why should one be interested in the mechanisms underlying the brain's response to targeted stimulation?
There are both theoretical and practical motivations.
Theoretically, despite the high-dimensionality of brain activity recordings, brain dynamics of spontaneous activity is known to be constrained to a lower-dimensional manifold \citep{shineLowDimensionalNeuralArchitecture2019, shineHumanCognitionInvolves2019, churchlandNeuralPopulationDynamics2012, chaudhuriIntrinsicAttractorManifold2019, cunninghamDimensionalityReductionLargescale2014, humphriesStrongWeakPrinciples2021}.
This implies the existence of a vast space of unexplored states that the brain does not typically access, unless perturbed with a sufficiently strong stimulus.
Targeted stimulation techniques enable us to access states that are not only inaccessible during spontaneous activity, but also difficult to access during sensory-evoked activity, hence initiating new forms of artificially evoked activity that transcend the brain's natural activation patterns.
Consequently, generative models present an important opportunity to test and calibrate mechanisms on datasets obtained through a unique experimental paradigm, potentially driving the discovery of new mechanisms of brain dynamics distinct from those that sufficiently capture spontaneous dynamics.
Practically, there is growing interest in developing clinical neuromodulation technologies using targeted stimulation to address a wide range of neurodegenerative and neurological disorders \citep{kurtinMovingPhenomenologicalPredictive2023}.
While trial-and-error methods for determining effective stimulation parameters have shown promising results for specific patients \citep{coleStanfordAcceleratedIntelligent2020, ramasubbuDosingElectricalParameters2018}, a wider deployment of neuromodulation necessitates a systematic approach that can more efficiently explore the vast stimulation parameter space and tailor the technology to each patient's needs \citep{wangChapterComputationalModeling2015, capogrossoComputationalOutlookNeurostimulation2020}.
An enhanced mechanistic understanding of the brain's response to the targeted stimulation techniques behind these technologies can therefore accelerate the development of more systematic personalized protocols, by guiding the interpretation of individual responses to different stimulation parameter combinations. 

In this review, we focus on the theoretically motivated opportunities in understanding the brain's response to targeted stimulation (for a review on the practically motivated opportunities mentioned above, see \citet{kurtinMovingPhenomenologicalPredictive2023, wangChapterComputationalModeling2015, tangColloquiumControlDynamics2018, acharyaBrainModelingControl2022} for examples).
With reference to a theoretical abstracted picture of brain dynamics, we argue for the existence of hidden nonlinear mechanisms of brain dynamics which are challenging to detect from spontaneous dynamics.
We highlight three available tools that we argue are essential to uncover these mechanisms:
(1) targeted stimulation techniques to make precise perturbations that can transition brain activity away from spontaneous patterns;
(2) measurement techniques to simultaneously record the brain's response to the perturbation as it reverts to spontaneous patterns; and
(3) generative models to test mechanisms that we hypothesize to capture the dynamics of recorded activity evoked by targeted stimulation.
Anchored around this argument, we explore in detail the opportunities that integrating these tools---that is, integrating different targeted stimulation and measurement techniques with different physiological, phenomenological, and data-driven models---provide in expanding our understanding of the various mechanisms underlying brain dynamics.

\section{A Dynamical Systems Framing of Targeted Brain Stimulation}

\begin{figure}[ht]
    \centering
    \includegraphics[width = \textwidth]{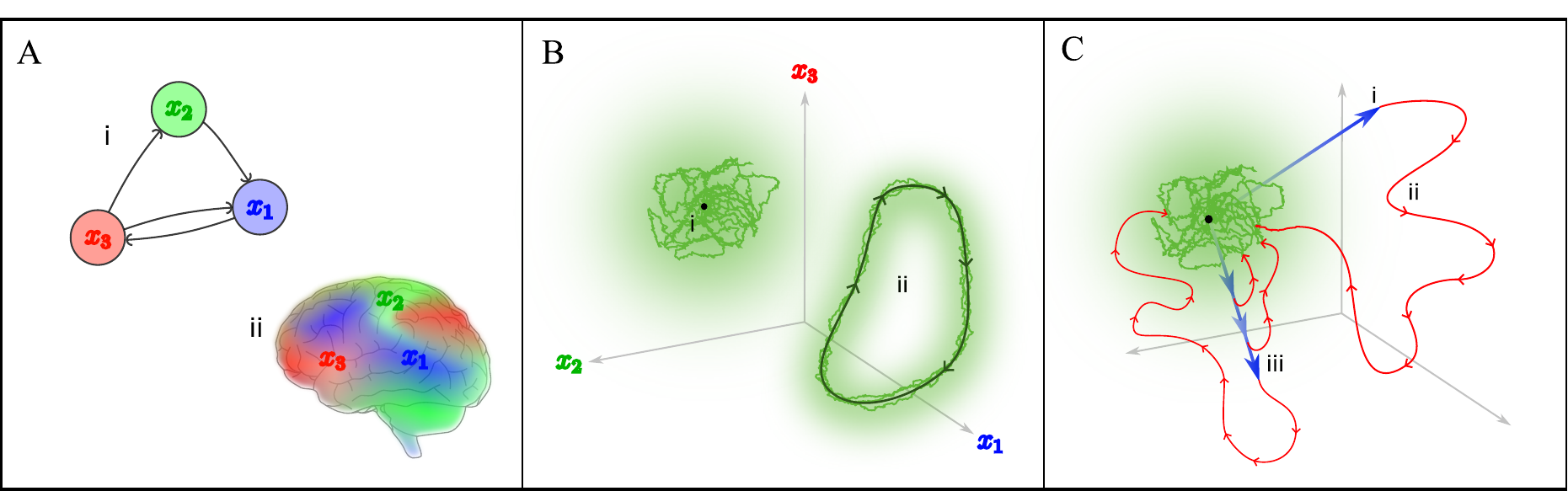}
    \caption{
    \textbf{An abstracted picture of brain dynamics that conceptualizes the dynamics measured from distinct experimental settings, while simplifying the complexities of brain activity measurements across other experimental conditions}.
    \textbf{A}: Brain activity dynamics can be represented as the evolution of a set of key state variables, such as: (i) the membrane potentials of individual neurons in a network; or (ii) activation of distributed modes of whole-brain activity.
    \textbf{B}: In this abstracted picture, spontaneous brain dynamics can be viewed as stochastic fluctuations about an attractor (a manifold that attracts nearby trajectories).
    Common attractor classes include fixed points \citep{vandenheuvelExploringBrainNetwork2010, sipCharacterizationRegionalDifferences2023}, labeled (i), and limit cycles \citep{shineHumanCognitionInvolves2019, shineLowDimensionalNeuralArchitecture2019}, labeled (ii).
    \textbf{C}: In contrast to spontaneous dynamics, stimulus-evoked brain dynamics consists of a perturbation, labeled (i), to a point that is typically far from the attractor, followed by a dynamic relaxation, labeled (ii), back to the attractor.
    Stimulus-evoked dynamics can exhibit interesting nonlinearities, where the response (red) qualitatively differs with the strength of the perturbation (blue), as depicted at (iii).
    }
    \label{fig:perturbationrelaxation}
\end{figure}

To concretely establish our theoretical motivation for understanding the brain's response to targeted stimulation, we present a geometric depiction of brain dynamics which abstracts the complexities of brain activity measurements across different spatial scales and species, shown in Fig.~\ref{fig:perturbationrelaxation}.
With reference to dynamical systems theory, we use this picture to argue that complex brain dynamics are likely underpinned by mechanisms that are not clearly apparent in the dynamics of commonly analyzed spontaneous activity, and that targeted stimulation techniques and measurement techniques are necessary tools to generate brain dynamics that manifest these hidden mechanisms. 
We also argue that the dynamics evoked by targeted stimulation are likely to be nonlinear, and that to rigorously test and validate the hidden mechanisms underlying these dynamics necessitates the use of generative modeling and simulation.

In this abstracted picture, the brain's activity is characterized as the evolution of a brain state over time.
Brain states can be encapsulated by a set of time-varying state variables, $\{x_1(t), x_2(t), x_3(t), \dots, x_N(t)\}$, and brain dynamics thus correspond to trajectories, $\mathbf{x}(t) = [x_1(t), x_2(t), \dots, x_N(t)]$, through the state space.
A specific spatial scale of interest can be addressed through appropriate selection of the state variables, from the activities of individual neurons in a population \citep{buzsakiLargescaleRecordingNeuronal2004} (depicted in Fig.~\ref{fig:perturbationrelaxation}A(i)), to the spatially distributed functional gradients or modes which describe patterns of population-scale activity over the whole brain \citep{huntenburgLargeScaleGradientsHuman2018, pangGeometricConstraintsHuman2023} (depicted in Fig.~\ref{fig:perturbationrelaxation}A(ii)).

By abstracting brain dynamics as trajectories through a state space, we can conceptualize the dynamics of brain activity measured from qualitatively distinct experimental settings.
Consider the commonly examined setting of spontaneous activity, which is measured when the brain is at rest and absent from any explicit external stimulus.
Measurements indicate that spontaneous brain states are concentrated at a subset of the state space with dimension much less than that of the state space, and do not deviate significantly from this subset \citep{shineLowDimensionalNeuralArchitecture2019, shineHumanCognitionInvolves2019, churchlandNeuralPopulationDynamics2012, chaudhuriIntrinsicAttractorManifold2019, cunninghamDimensionalityReductionLargescale2014}.
Consistent with these properties, spontaneous dynamics in our abstracted picture can be conceptualized as trajectories concentrated near \textit{attractors}, which are low-dimensional subsets of the state space that attract neighboring trajectories \citep{strogatz2018nonlinear}.
Additional sensory inputs and background biological processes can also cause trajectories to temporarily deviate from and return to the attractor, effectively yielding stochastic fluctuations about the attractor.
Attractors, which define spontaneous dynamics across different experimental conditions, can be further categorized into classes of similar geometric features; such as stable fixed points \citep{vandenheuvelExploringBrainNetwork2010, sipCharacterizationRegionalDifferences2023} (depicted in Fig.~\ref{fig:perturbationrelaxation}B(i)), or flows along a more complex low-dimensional manifold such as a limit cycle \citep{shineHumanCognitionInvolves2019, shineLowDimensionalNeuralArchitecture2019, churchlandNeuralPopulationDynamics2012} (depicted in Fig.~\ref{fig:perturbationrelaxation}B(ii)).
Through classes of attractors, we can concisely capture the intrinsic dynamical properties of spontaneous brain activity in this abstracted picture, whilst aggregating the impact of other experimental conditions.

In addition to the low-dimensional attractor where trajectories of spontaneous dynamics are concentrated, are a myriad of possible trajectories of brain dynamics through states far from the attractor.
Unfortunately, as brain states move further from the attractor, they become less likely to be observed in spontaneous settings.
However, our abstracted picture can conceptualize an alternative \textit{stimulus-evoked} setting that employs external intervention to reach these distant states, as shown in Fig.~\ref{fig:perturbationrelaxation}C.
Initially positioned near the attractor (depicted here as a fixed point), the brain is first perturbed with a sufficiently strong stimulus to a distant state (Fig.~\ref{fig:perturbationrelaxation}C(i)), and subsequently relaxes back to the attractor following stimulus termination (Fig.~\ref{fig:perturbationrelaxation}C(ii)).
According to nonlinear dynamical systems theory, a central characteristic of stimulus-evoked dynamics is the nonlinear nature of the relaxation trajectory \citep{strogatz2018nonlinear}, where its shape is expected to be highly sensitive to the magnitude of the perturbation, as illustrated in Fig.~\ref{fig:perturbationrelaxation}C(iii).
However, while such nonlinearities can produce a diverse range of trajectories far from the attractor, their effects tend to vanish as the trajectory approaches the attractor governing spontaneous dynamics, implying that spontaneous dynamics can be sufficiently approximated by simpler linear models.
Generative modeling studies have indeed shown that linear models can capture properties of macroscale spontaneous brain dynamics as effectively as nonlinear models, such as the established resting state functional connectivity benchmark  \citep{nozariMacroscopicRestingstateBrain2023, messePredictingFunctionalConnectivity2015, hosakaLinearModelsReplicate2024}. 
These findings, supported by predictions of nonlinear dynamical systems theory, suggest that the nonlinear mechanisms shaping stimulus-evoked brain dynamics in this dynamical systems picture are difficult to be detected in spontaneous settings.
Thus, to uncover and understand these nuanced hidden mechanisms of stimulus-evoked dynamics in practice (beyond this abstracted picture), we require an experimental paradigm which allows us to perturb the brain away from its spontaneous attractor (Fig.~\ref{fig:perturbationrelaxation}C(i)) and resolve the nonlinear relaxation trajectory (Fig.~\ref{fig:perturbationrelaxation}C(ii)), alongside a model which allows us to question and validate the nonlinear mechanisms which underpin these trajectories.

To this end, we bring attention to a practically feasible strategy that integrates targeted stimulation techniques and measurement techniques with generative models.
Targeted stimulation techniques can perturb the brain from the attractor to distant states, with spatial coverage and precision that surpass traditional sensory stimuli \citep{deisserothOptogenetics10Years2015, kellerCorticocorticalEvokedPotentials2014, rogaschAssessingCorticalNetwork2012, rothDREADDsNeuroscientists2016}.
High precision selection of states is crucial, as the nonlinear dynamics of stimulus-evoked activity would be highly sensitive to perturbation strengths.
Compatible measurement techniques play an indispensable role, by simultaneously tracking the brain’s relaxation trajectory back to the attractor at sufficient spatial and temporal resolution.
Finally, the simulation approach of generative models is essential for linking hypothesized mechanisms to the dynamics observed in measurements; as a mechanistic understanding of nonlinear properties is challenging to pursue from static statistical measures alone \citep{johnItTimeLinking2022}.

Targeted stimulation techniques, measurement techniques, and generative models are therefore three important tools which, when integrated, can uncover and provide understanding of new mechanisms of brain dynamics that are concealed in spontaneous settings. 
By conducting experiments with stimulation and measurement techniques, we can generate datasets of stimulus-evoked dynamics with potentially new forms of nonlinearities.
By constraining generative models with datasets from these experiments, we can rigorously test and validate hypotheses about the novel mechanisms which drive their nonlinear dynamics. 
To explore the opportunities that emerge from this integration, we first delve into the different experiments that can be executed.

\section{Targeted Stimulation Experiments}

Through an abstracted picture of brain dynamics, illustrated in Fig.~\ref{fig:perturbationrelaxation}, we have argued why targeted stimulation experiments play an important role in revealing novel mechanisms of brain dynamics in measured brain activity.
This picture serves another purpose: it describes the fundamental dynamical properties of activity in any targeted stimulation experiment, while abstracting away details of the specific techniques used.
This interpretation bridges a wide body of stimulation and measurement techniques developed from disjoint fields, each with their own mechanics of action, spatial scale, and species compatibility.
Here, we review a diverse range of key stimulation and measurement techniques that can be used in a targeted stimulation experiment, contextualizing their features within our unified picture.

\begin{figure}[ht]
    \centering
    \includegraphics[width = \textwidth]{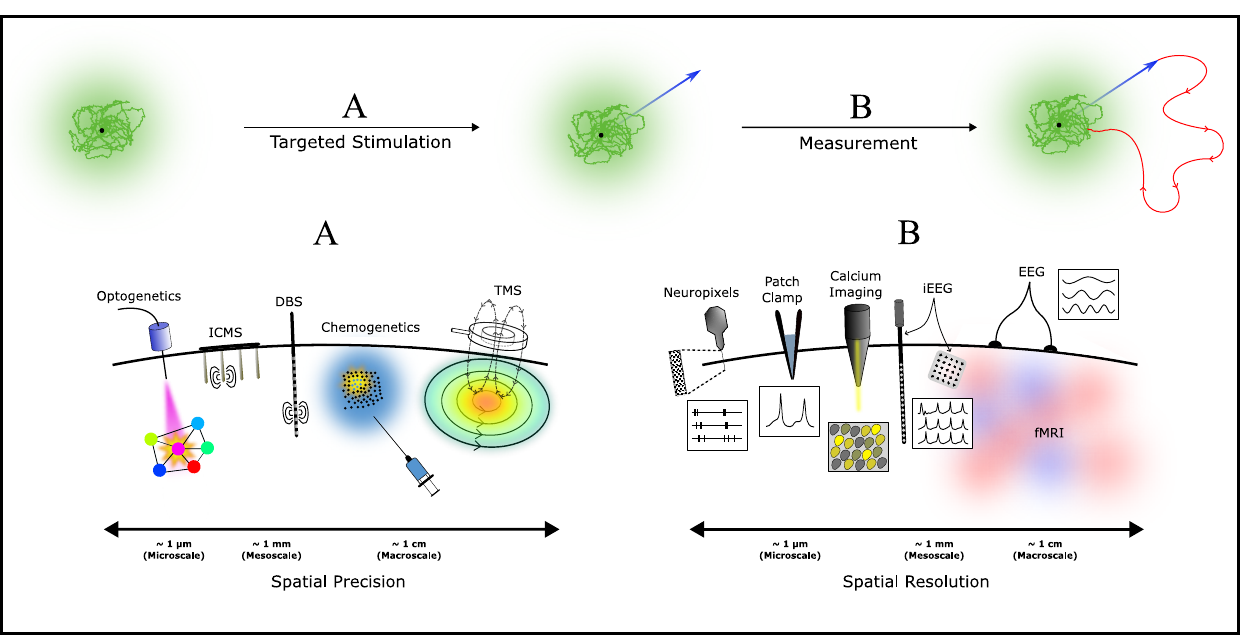}
    \caption{
    \textbf{Stimulus-evoked brain activity can be generated through a diverse range of targeted stimulation techniques and measured by a range of measurement techniques at different spatial scales.}
    \textbf{A}: Targeted stimulation techniques can perturb the brain to novel brain states at different degrees of spatial precision.
    From neuron-scale techniques, such as optogenetics; to mesoscale techniques, such as ICMS and DBS; to macroscale techniques, such as chemogenetics and TMS. 
    \textbf{B}: Different measurement techniques can measure the brain's response to targeted stimulation at different spatial resolutions.
    From neuron-resolution modalities, such as Neuropixels electrodes, patch clamps, and calcium imaging; to mesoscale modalities, such as iEEG systems; to macroscale modalities, such as EEG and fMRI.
    }
    \label{fig:methods}
\end{figure}

In a targeted stimulation experiment, a stimulation technique perturbs the brain to a state distant from the attractor representing spontaneous dynamics (Fig.~\ref{fig:perturbationrelaxation}(i)), and a measurement technique captures the brain's subsequent relaxation trajectory back to the attractor (Fig.~\ref{fig:perturbationrelaxation}(ii)).
Key targeted stimulation and measurement techniques are illustrated in Fig.~\ref{fig:methods}.
Based on the spatial scale in which brain states are defined, it is crucial to select an appropriate targeted stimulation technique to elicit a precise perturbation in this state space, and an appropriate measurement technique to accurately capture the relaxation.

\subsection{Targeted Stimulation techniques}

Different targeted stimulation techniques can be selected to elicit perturbations at a spatial precision of choice.
Microscale techniques can precisely stimulate individual neurons \citep{deisserothOptogenetics10Years2015}, while meso-to-macroscale techniques can stimulate neural populations ranging from millimeters to centimeters in size \citep{flesherIntracorticalMicrostimulationHuman2016, kringelbachTranslationalPrinciplesDeep2007, rogaschAssessingCorticalNetwork2012}.
Figure~\ref{fig:methods}A illustrates key targeted stimulation techniques, ordered by degrees of spatial precision.
We review these techniques below.

At the microscale, techniques can stimulate a single target neuron or a set of neurons.
The most prevalent microscale stimulation technique is optogenetics, in which target neurons are genetically modified with injected photoreceptor proteins called opsins, and later exposed to light of a specified wavelength to generate action potentials \citep{deisserothOptogenetics10Years2015}.
Two-photon microscopy allows the light beam to be made sufficiently narrow to be directed exclusively to the target neurons \citep{rickgauerTwophotonExcitationChannelrhodopsin22009, tongSingleCellVivo2023, shemeshTemporallyPreciseSinglecellresolution2017}.

At the mesoscale, neural populations can be similarly stimulated using optogenetics, or via electrical methods such as intracortical microstimulation (ICMS) and deep brain stimulation (DBS).
A compelling benefit of using optogenetics at this scale is its ability to target specific cell-types, cortical layers, and projection targets through viral-vector gene-delivery techniques \citep{yizharOptogeneticsNeuralSystems2011}.
The alternative methods, ICMS and DBS, are electrical techniques which stimulate local populations in the vicinity of an implanted electrode using electric fields \citep{flesherIntracorticalMicrostimulationHuman2016, kringelbachTranslationalPrinciplesDeep2007}. 
While these methods lack the forms of specificity that can be achieved with optogenetics, they can be applied in human brains as it does not require genetic modification of the target neural tissue.
Local populations near the cortical surface can be targeted by ICMS through a chronically implanted multielectrode array \citep{flesherIntracorticalMicrostimulationHuman2016}, whereas populations of varying depth down to subcortical structures can also be targeted by DBS through electrodes implanted deep in the brain \citep{kringelbachTranslationalPrinciplesDeep2007}.

Finally, macroscale neural populations can be stimulated using electrical methods such as transcranial magnetic stimulation (TMS), or chemogenetics which is a genetic modification technique similar to optogenetics.
TMS uses an external electromagnetic coil to generate a magnetic field, which then induces an electric field over the target population \citep{rogaschAssessingCorticalNetwork2012, tremblayClinicalUtilityProspective2019}.
TMS is a non-invasive technique, as the generated magnetic field passes across the scalp and skull, however the difficulty in downsizing the coil has mainly restricted its use to primate brains \citep{alekseichukComparativeModelingTranscranial2019}.
In smaller rodent brains, macroscale populations can be stimulated using chemogenetics, in which the target population is injected with an engineered protein called Designer Receptors Exclusively Activated by Designer Drugs (DREADDs), and consequently activated by an orally administered designer drug \citep{rothDREADDsNeuroscientists2016}.
Similar to optogenetics, chemogenetics benefits from the forms of specificity provided by viral-vector techniques, but it is able to stimulate much larger neural populations, as the designer drug diffuses throughout the brain when orally administered \citep{sternsonChemogeneticToolsInterrogate2014}.
Chemogenetic stimulations also have the unique ability to modulate brain activity for sustained periods ranging from hours to weeks \citep{muirWiringDepressedBrain2019}.

\subsection{Measurement Techniques}

Following the perturbation elicited by the selected targeted stimulation technique, we can simultaneously measure the brain's relaxation trajectory back to the attractor of spontaneous activity.
Collections of such measurements form a dataset of stimulus-evoked activity, which we have argued to reveal undiscovered mechanisms of brain dynamics upon further analysis.
To capture the relaxation, the selected measurement technique must measure the exhibited brain activity at a matching spatial resolution.
Microscale measurement techniques can resolve the response from individual neurons; while mesoscale and macroscale measurement techniques capture aggregated activity of neural populations.
Figure~\ref{fig:methods}B illustrates key measurement techniques, ordered by degrees of spatial resolution.
We review these techniques below.

At a microscale resolution, single-unit recording methods capture the spiking activity of individual neurons.
The classic single-unit recording method is whole-cell patch-clamp electrophysiology, which reads the intracellular potential from an electrode of a micropipette tightly suctioned to the target neuron membrane \citep{hillIntroductionPatchClamp2021}.
The spatial coverage of patch clamps is however limited to a few neurons.
An alternate emerging modality is calcium imaging, which measures changes in levels in calcium concentration as an indirect proxy for spiking activity \citep{grienbergerImagingCalciumNeurons2012}.
Modern microscopy techniques can measure calcium levels of hundreds of neurons, recently measuring activity across all neurons in \textit{C. elegans} in response to neuron scale optogenetics
\citep{randiNeuralSignalPropagation2023}.
With emerging microelectrode arrays, such as the recently developed Neuropixels probes \citep{steinmetzNeuropixelsMiniaturizedHighdensity2021}, it is now possible to directly measure neuronal spiking with a spatial coverage comparable to that of calcium imaging.

Activity of a mesoscale population is conventionally measured by the local field potential (LFP) of its extracellular medium, which aggregates the spiking of its constituent neurons \citep{buzsakiOriginExtracellularFields2012}.
LFPs can be recorded by intracranial electroencephalographic (iEEG) electrodes penetrating the cortex.
Commonly used iEEG electrode designs include subdural strip or grid electrodes, which can monitor LFPs from multiple positions from the cortical surface, or depth electrodes, which can measure at multiple depths at a single surface position \citep{lachauxIntracranialEEGHuman2003}.

Activity of a macroscale neural population could be measured by simultaneously tracking mesoscale populations with multiple iEEG systems, however this option can only be implemented in rare surgical scenarios (for example, the F-TRACT project \citep{jedynakVariabilitySinglePulse2023}), as many electrodes become excessively invasive.
Instead, there exist non-invasive functional neuroimaging techniques which make proxy measurements at this spatial scale.
Electroencephalography (EEG) can measure electrical potentials from electrodes attached around the scalp at high temporal resolution, making it suitable for measuring transient responses to stimulation such as TMS \citep{rogaschAssessingCorticalNetwork2012}.
Complementing EEG is functional magnetic resonance imaging (fMRI), which measures the blood oxygen level dependent (BOLD) signals tracking underlying neural activity.
Despite its limited temporal resolution, fMRI with its high spatial coverage can be used to measure responses to a macroscale modulatory stimulation method such as chemogenetics \citep{zerbiRapidReconfigurationFunctional2019, markicevicCorticalExcitationInhibition2020, markicevicNeuromodulationStriatalD12023}.

In summary, there are a range of sophisticated targeted stimulation techniques that can perturb precise neurons or broad neural populations; and measurement techniques which can simultaneously resolve the subsequent relaxation of neuronal spiking activity or aggregated neural population activity.
Thanks to these technological advancements, datasets of stimulus-evoked activity can be generated through targeted stimulation experiments at any spatial scale of our interest: micro-, meso-, or macroscale.
As argued through an abstracted picture of brain dynamics (Fig.~\ref{fig:perturbationrelaxation}), stimulus-evoked brain activity may manifest highly nonlinear dynamics, driven by novel mechanisms.
However, integrating these experiments with generative models allows us to rigorously test hypotheses about these nonlinear mechanisms at any spatial scale.
In the next section, we detail how different generative models can be utilized to harness this significant opportunity.

\section{Generative Models of Brain Activity}

Generative models provide a mechanistic account of the spatiotemporal dynamics of brain activity observed under any experimental setting. 
Mechanisms are encoded as a set of dynamical rules, such as difference equations or differential equations, which govern the evolution of brain activity over time. 
By simulating these governing rules, we can generate a spatially distributed time series of brain activity, which can be compared against activity from experimental data. 
Mechanisms of generative models can also be designed to generate nonlinear time series, allowing us to capture the nonlinear dynamical properties of stimulus-evoked activity.
Aside from the ability to generate neuronal dynamics, the hypothesized mechanisms of generative models can also be assessed and validated by integrating the model with a target experimental dataset. 
This process may involve model selection and model calibration, where the model's parameters and possible structural characteristics are optimized to explain the dataset as accurately as possible \citep{ramezanian-panahiGenerativeModelsBrain2022}.
One can additionally assess the model's predictions of the target dynamics through an uncertainty quantification phase, in which statistical methods quantify the reliability of model's parameter estimates and predictions \citep{jhaFullyBayesianEstimation2022, penasParameterEstimationWholebrain2024}, and sensitivity analyses evaluate how robust the output dynamics are to small changes in the inferred parameters  \citep{hashemiSimulationbasedInferenceVirtual2024, baldyInferenceMacroscopicDynamics2024}.
Through the generation of nonlinear dynamics, and the integration with targeted stimulation datasets, generative models are therefore indispensable tools in the endeavor to infer mechanisms of brain dynamics revealed in the activity evoked by targeted stimulation.

\begin{figure}[ht]
    \centering
    \includegraphics[width = \textwidth]{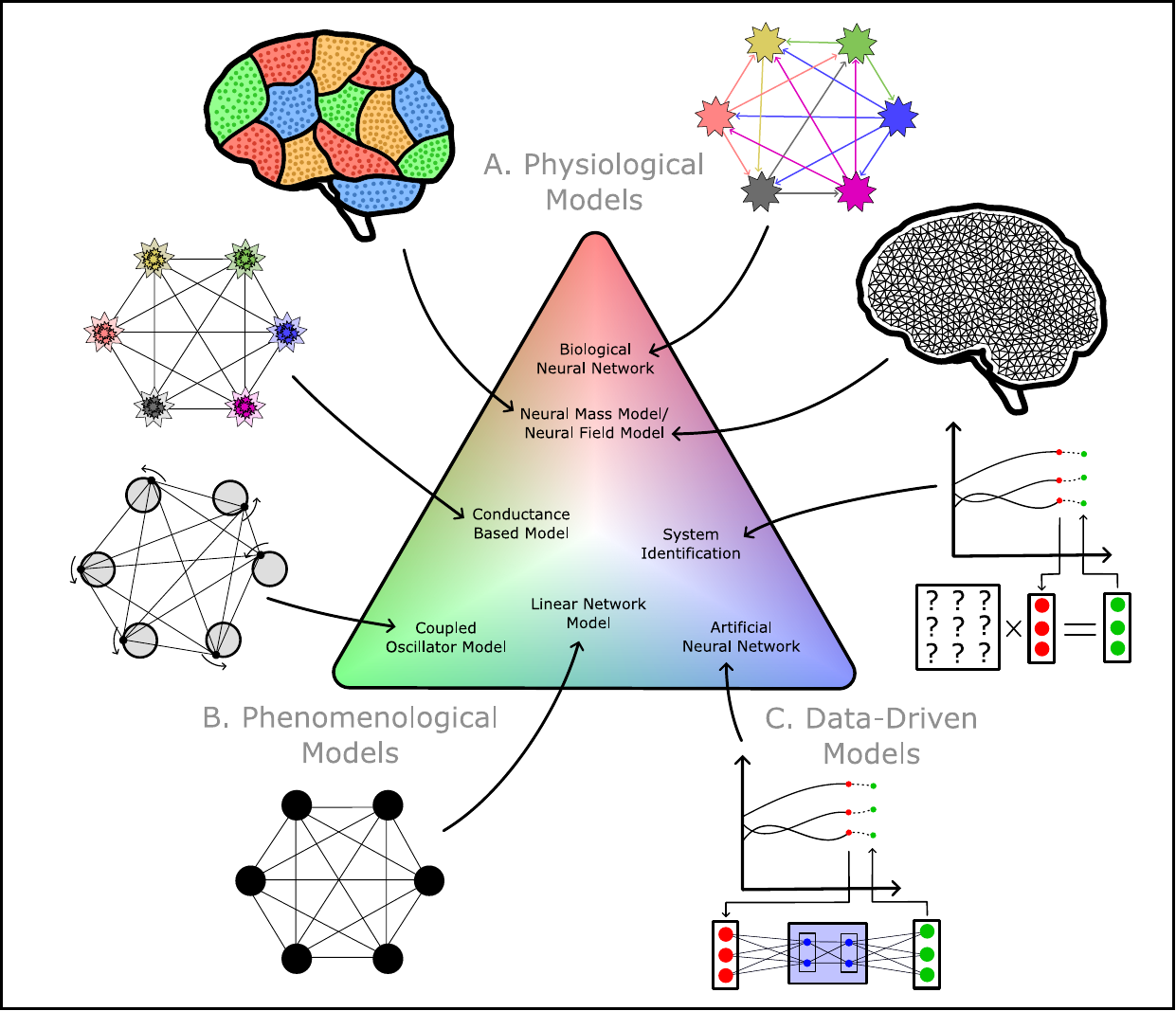}
    \caption{\textbf{Generative models come in a variety of flavors, and can be positioned along a spectrum between three extremes, depending on the type of mechanistic account of brain dynamics they provide}.
    \textbf{A}: Physiological models---such as biological neural networks, neural mass models, and neural field models---model physiological processes of brain activity with physically measurable biological quantities \citep{breakspearDynamicModelsLargescale2017, decoDynamicBrainSpiking2008}.
    \textbf{B}: Phenomenological models---such as linear network models, coupled oscillator models, and conductance-based models---model abstracted mechanisms that can capture essential dynamical properties of brain activity with minimal physiological detail \citep{cabralFunctionalConnectivityDynamically2017, kimLinearDynamicsControl2020}.
    \textbf{C}: Data-driven models---such as models constructed through system identification or artificial neural networks---learn statistical associations of brain dynamics from sample data to make accurate predictions of brain dynamics \citep{acharyaBrainModelingControl2022, ramezanian-panahiGenerativeModelsBrain2022}.}
    \label{fig:models}
\end{figure}

Generative models encode various forms of mechanisms to provide different forms of accounts of brain dynamics.
Figure~\ref{fig:models} illustrates a spectrum of key generative models, organized by the type of mechanisms they encode.
By choosing and employing an appropriate model, we can test and validate the novel mechanisms that we hypothesize to drive the dynamics evoked by targeted stimulation.
\textit{Physiological models}, for example, use physically measurable quantities to quantitatively encapsulate complex physiological processes in the brain  \citep{breakspearDynamicModelsLargescale2017, decoDynamicBrainSpiking2008}.
In contrast, \textit{phenomenological models} use mathematical abstractions to capture the core spatiotemporal properties of brain dynamics hypothesized by the mechanism \citep{cabralFunctionalConnectivityDynamically2017, kimLinearDynamicsControl2020}.
Finally, we distinguish \textit{data-driven models}, which learn statistical associations from sample data to make accurate predictions of brain dynamics \citep{acharyaBrainModelingControl2022, ramezanian-panahiGenerativeModelsBrain2022}.
Despite their varying mechanisms, all generative models in Fig.~\ref{fig:models} can simulate both spontaneous and stimulus-evoked activity as abstracted by the dynamical systems picture in Fig.~\ref{fig:perturbationrelaxation}.
The relevant output variables construct a state space where brain activity follows a trajectory of brain states (Fig.~\ref{fig:perturbationrelaxation}A).
Introducing stochastic noise as an input variable concentrates the trajectory around an attractor, reflecting the model's depiction of spontaneous dynamics at the corresponding spatial scale (Fig.~\ref{fig:perturbationrelaxation}B).
Contrastingly, setting the input variable to a deterministic spatiotemporal profile aligned with the experimental stimulus would perturb the brain state away from the attractor before drawing it back via the relaxation trajectory (Fig.~\ref{fig:perturbationrelaxation}C). 

In the subsections below, we review how physiological, phenomenological, and data-driven models are constructed, and how the specific types of mechanisms they encode can contribute to our understanding of the dynamics evoked by targeted stimulation.
For each model category, we provide examples of modeling studies which have tested mechanisms underlying brain dynamics through an integration with datasets of stimulus-evoked activity, with a primary focus on emerging datasets of targeted stimulation.
We also highlight how findings of some studies support our core argument presented through the abstracted picture in Fig.~\ref{fig:perturbationrelaxation}: that stimulus-evoked dynamics could be driven by novel mechanisms of brain dynamics which play a more peripheral role in shaping spontaneous dynamics.

\newpage

\subsection{Physiological Models}

In physiological models, variables and parameters represent physical quantities such as cell-body potentials or synaptic coupling strengths, and interact in accordance to experimentally supported physiological processes \citep{maReviewDynamicsNeuron2017, breakspearDynamicModelsLargescale2017, decoDynamicBrainSpiking2008}.
Physically measurable variables and parameters assist in interpreting the functional significance of individual variables, and the effects of changes in static parameters on the resultant dynamics.
Crucially, the physiological formulation allows predictions of the model to be tested not just against the model's ability to capture the predicted brain dynamics, but also against physical measurements from physiological experiments.
Physiological models can study the dynamics of responses to a wide range of targeted stimulation techniques, with input variables quantifying the technique's specific mechanics of action.
For example, a physiological model could be formulated to directly compare responses to optogenetics, which directly depolarizes neuron membranes, with responses to electrode stimulation, which feeds a current into the extracellular medium.
We can construct physiological models to test mechanisms of physiological processes at different spatial scales, ranging from the microscale to whole brain scale.

To capture microscale brain dynamics, we can use \textit{biological neural networks} which simulate the activity of interconnected neurons \citep{maReviewDynamicsNeuron2017}.
Separate single-neuron models, such as the Hodgkin--Huxley model \citep{hodgkinQuantitativeDescriptionMembrane1952}, are coupled to capture neuron--neuron interactions reflecting underlying synaptic processes.
By assigning separate models to each neuron, biological neural networks can simulate dynamics with realistic heterogeneity among neurons.
Specific parameters can be assigned to heterogeneous classes of neurons, such as excitatory and inhibitory neurons \citep{gastNeuralHeterogeneityControls2024, stefanescuLowDimensionalDescription2008, landauImpactStructuralHeterogeneity2016, arkhipovVisualPhysiologyLayer2018, billehSystematicIntegrationStructural2020}, or classes of intricate morphological features \citep{aberraBiophysicallyRealisticNeuron2018, aberraSimulationTranscranialMagnetic2020}.
Ongoing projects to create expansive classification databases such as the Blue Brain Project \citep{markramBlueBrainProject2006} and the Allen Cell Types Database \citep{gouwensSystematicGenerationBiophysically2018}, alongside technological advances in computing power, suggest that future biological neural networks could simulate brain dynamics at neuronal resolution whilst spanning large spatial scales.

Zooming out from the microscale, mean-field population models instead track the mean activity of all neurons in a population, without tracking each constituent neuron \citep{decoDynamicBrainSpiking2008, breakspearDynamicModelsLargescale2017}.
Crucially, this mean-field approach to model construction requires assumptions on the connectivities between populations in space.
For example, there are \textit{neural field models} \citep{jirsaFieldTheoryElectromagnetic1996, jirsaDerivationMacroscopicField1997}, which assume isotropic connectivity between populations to allow for efficient computation of population-scale activity across continuous space \citep{robinsonPropagationStabilityWaves1997, robinsonDynamicsLargescaleBrain2002, robinsonPropagatorTheoryBrain2005, decoDynamicBrainSpiking2008}.
Another class of mean-field models across populations are \textit{neural mass models} or brain-network models \citep{pathakWholeBrainNetworkModels2022, breakspearDynamicModelsLargescale2017}, which coarse-grain neural populations into discrete spatial regions, and then couple these regions to one another in a network, using measured structural connectome data \citep{chaudhuriLargeScaleCircuitMechanism2015, decoKeyRoleCoupling2009, decoDynamicalConsequencesRegional2021}.
Notwithstanding the limitations in accuracy and physical realism due to mean-field assumptions \citep{robinsonPhysicalBrainConnectomics2019}, the significant reduction in the number of variables allows mean-field population models to be computed much more efficiently than the equivalent hypothetical biological neural network of the same spatial scale.

By designing stimulation inputs with unique spatiotemporal profiles, physiological models integrated with targeted stimulation datasets can help investigate the mechanisms behind varying responses to different stimulation forms. 
For instance, a biological neural network of the rat somatosensory cortex revealed how morphological features like axon length shape neuron activation under ICMS and TMS under various stimulation parameters \citep{aberraBiophysicallyRealisticNeuron2018}.
Another biological neural network of the mouse and monkey somatosensory cortex investigated recurrent mechanisms underpinning the difference in responses between optogenetic and visual sensory stimulation \citep{sanzeniMechanismsUnderlyingReshuffling2023}.
Additionally, physiological models can account for nonlinearities observed in experimental responses to targeted stimulation.
For example, a biological neural network of the macaque motor cortex demonstrated potential nonlinear mechanisms underpinning distortive effects of ICMS on the dynamics of ongoing task-related activity \citep{osheaDirectNeuralPerturbations2022}.
At the macroscale, a neural field model with nonlinear synaptic mechanisms studied the calcium-dependent mechanisms underlying the nonlinear effects of TMS-induced plasticity \citep{fungNeuralFieldTheory2013}. 
A similar neural field model demonstrated significant differences in local synaptic strength parameters when fitted to spontaneous EEG versus evoked potentials \citep{kerrPhysiologybasedModelingCortical2008}.
This discrepancy was reconciled with a dynamic synaptic strength governed by a nonlinear gain modulation mechanism \citep{babaie-janvierNeuralFieldTheory2020}.

\subsection{Phenomenological Models}

While physiological models offer realistic biophysically constrained explanations of how brain activity stems from independently measurable physiological processes, it is also useful to avoid excessive physiological details that complicate the model's dimensionality and human interpretability \citep{robinsonTenRulesEffective2022}.
An alternative approach, taken by phenomenological models, is to replace a full physiological description with simpler canonical dynamical systems, which preserve the dynamics hypothesized to be exhibited by the underlying complex physiological processes \citep{cabralFunctionalConnectivityDynamically2017, kimLinearDynamicsControl2020}.
This abstraction is supported by the observation that many physiologically realistic models of stable brain dynamics exhibit canonical dynamical structures such as fixed points and limit cycles \citep{ramezanian-panahiGenerativeModelsBrain2022, freyerBiophysicalMechanismsMultistability2011, siuExtractingDynamicalUnderstanding2022}.
While the variables and parameters of phenomenological models are not physiologically interpretable (and thus not experimentally measurable), they simplify the model's complexity and the systems-level interpretation of the resulting simulated dynamics.
Phenomenological models are thus valuable for investigating the core spatially distributed properties of the dynamics of networks of neurons or brain regions in response to targeted stimulation, without comprehensive detail of the local physiological mechanisms that underpin each neuron or brain region's dynamics.

Phenomenological models vary in the sophistication of dynamics exhibited by their canonical structures.
A simple example is a \textit{linear network model}, which uses a linear system of equations to model the dynamics of a network of nodes representing neurons or neural populations \citep{kimLinearDynamicsControl2020, tangColloquiumControlDynamics2018}.
Each node's dynamics are driven by a linear combination of activities from coupled nodes, and any external perturbation such as targeted brain stimulation.
Linear models are a means to capture correlations between the activity of coupled neurons or neural populations with the simplest form of node-to-node coupling interactions \citep{abdelnourNetworkDiffusionAccurately2014, rajNetworkDiffusionModel2012}.
They may also allow for closed-form expressions for network responses to inputs, mathematically linking the system dynamics directly to its underlying coupling structure \citep{parkesAsymmetricSignalingHierarchy2022, guOptimalTrajectoriesBrain2017}.  

Generalizing the linear setting, there are nonlinear network models governed by nonlinear systems of equations, which can capture more sophisticated forms of local dynamics.
A simple example is linear threshold models \citep{misicCooperativeCompetitiveSpreading2015}, which extend linear models by incorporating an additional threshold to the linear combination input.
More sophisticated \textit{coupled oscillator models} aim to capture oscillatory properties of brain dynamics as observed in recordings of brain activity across spatial scales \citep{breakspearGenerativeModelsCortical2010, decoDynamicsRestingFluctuations2017}.
Commonly used oscillators include the Kuramoto oscillator which oscillates with constant amplitude \citep{golloMappingHowLocal2017, cabralRoleLocalNetwork2011}, and the Stuart--Landau (Hopf) oscillator which can also change in amplitude \citep{decoDynamicsRestingFluctuations2017, decoPerturbationWholebrainDynamics2018, decoSingleMultipleFrequency2017}.
A more physiologically inspired nonlinear model is a \textit{conductance-based model}, which models population-level activity abstracted as a simpler biological neural network \citep{breakspearModulationExcitatorySynaptic2003, larterCoupledOrdinaryDifferential1999, honeyPredictingHumanRestingstate2009}.
Here, the aggregate activity of each population is tracked by the dynamics of a single neuron model, which assumes a sufficiently strong coherence between constituent neurons of each population \citep{breakspearDynamicModelsLargescale2017}.
Despite this strong physiological assumptions, conductance-based models are capable of exhibiting population-scale spatiotemporal dynamics with interesting properties that are commonly observed in biological neural networks.

In comparison to physiological models, the abstractions incorporated in the local dynamics of phenomenological models limit their ability to distinguish between the dynamics evoked by different targeted stimulation techniques.  
However, phenomenological models can be used to more directly investigate how the brain's underlying structure shapes the spatially distributed properties of its stimulus-evoked dynamics under an arbitrary stimulus input, as they can isolate the effects of complex local dynamics.
For example, a coupled oscillator model integrated with mouse optogenetic stimulation data demonstrated how heterogeneous long-range structural connections allow stimulus-evoked activity to reveal a rich repertoire of novel dynamical responsive functional networks, extending the established repertoire of resting state networks observed in spontaneous dynamics \citep{spieglerSelectiveActivationRestingState2016, spieglerSilicoExplorationMouse2020}.
Phenomenological models can also investigate the extent to which deviations in spatiotemporal dynamics between stimulus-evoked and spontaneous activity are influenced by the stimulus position. 
For example, a coupled oscillator model demonstrated how TMS-induced changes in functional connectivity varied between stimulating highly connected hub regions in hub regions and stimulating weakly connected peripheral regions \citep{golloMappingHowLocal2017}.  
Another coupled oscillator model studied how the influence of stimulus position varied between different global brain states such as wakefulness and deep sleep \citep{decoPerturbationWholebrainDynamics2018}.

\subsection{Data-Driven Models}

The governing rules of phenomenological and physiological models summarized above reflect qualitatively interpretable mechanisms underlying brain dynamics, such as the internal physiological processes of neurons in biological neural networks, or the oscillatory behaviors of population-scale neural recordings in a whole-brain oscillator model.
In contrast, data-driven models encode statistical mechanisms which can describe the quantitative structure of the governing rules without a qualitative interpretation \citep{acharyaBrainModelingControl2022, ramezanian-panahiGenerativeModelsBrain2022}.
Through the use of statistical techniques, data-driven models are capable of exploring a wide range of possible governing rules which reflect encoded statistical mechanisms, enabling highly accurate predictions of brain dynamics that are unconstrained by predefined physiological or phenomenological mechanisms.
This flexibility is particularly advantageous when studying responses to targeted stimulation, where underlying qualitative mechanisms may be difficult to hypothesize in advance.
At the same time, choosing rule structures requires careful judgment: models need enough coefficients to predict complex dynamics, but not too many under-constrained coefficients which undermine the model's generalization ability.

The range of different possible statistical properties of brain activity that a data-driven model can capture, varies with the range of governing rules that it can explore.
For example, there are data-driven models which use the process of \textit{system identification} to infer a set of mathematical equations that govern the dynamics of brain activity from the dataset \citep{acharyaBrainModelingControl2022}.
While system identification algorithms are limited to exploring and discovering equation-based governing rules, these equations enable the interpretation of the learned statistical properties of individual time series, and pairwise dependencies between time series.
Commonly used system identification procedures include linear time series models which learn simple linear equation structures \citep{ljungSystemIdentification1998}; and procedures which can estimate nonlinear equation structures \citep{billings2013nonlinear, bruntonDiscoveringGoverningEquations2016, williamsDataDrivenApproximation2015}.
Comparing different system identification procedures therefore provides an opportunity to detect nonlinearities in stimulus-evoked dynamics, by comparing the capacities of linear and nonlinear equations in predicting responses \citep{acharyaPredictiveModelingEvoked2023, yangModellingPredictionDynamic2021, changMultivariateAutoregressiveModels2012}.

There are also more flexible data-driven models such as \textit{artificial neural networks} (not to be confused with biological neural networks), which can be used to solve a more general problem of learning a predictive mapping between an input and output variable from a dataset of observations \citep{acharyaBrainModelingControl2022}.
The majority of artificial neural network architectures, including the simplest feed-forward architecture, learn nonlinear mappings between variables as a sequence of layers of units, including the input layer, output layer, and hidden layers \citep{ljungDeepLearningSystem2020}.
Each unit receives inputs from units of the previous layer, then feeds the input into an activation rule before sending it to units of the next layer.
While less interpretable than the systems of equations discovered by system identification methods, artificial neural networks are more flexible in replicating a wider range of governing rule structures, including those with high orders of nonlinearities that would otherwise be computationally expensive to be captured as a set of equations \citep{hornikMultilayerFeedforwardNetworks1989, vyasComputationNeuralPopulation2020}.
This flexibility potentially enables artificial neural networks to create future personalized models that make accurate predictions of individual responses to targeted stimulation, adapting important individual-specific structural and functional characteristics which are difficult to encode as physiological or abstracted mechanisms \citep{misraLearningBrainDynamics2021}.

A systematic comparison of data-driven models can be used to test whether hypothesized statistical mechanisms are necessary in capturing responses to targeted stimulation.
For example, a comparative study of responses to DBS found significant differences in fitted ARX model parameters between spontaneous and stimulus-evoked dynamics, thus suggested the use of a switched linear model as a simple nonlinear model that switches between different parameters with and without stimulation \citep{acharyaPredictiveModelingEvoked2023}.
Similar predictive powers were also found between switched linear models and less interpretable feedforward neural networks, demonstrating that switched linear models sufficiently explained responses to DBS without significantly trading off predictive capabilities \citep{acharyaPredictiveModelingEvoked2023}.
Another avenue of systematic comparison involves an architecturally distinct family of recurrent neural networks, which use recurrent connections to accumulate memory of an indefinite number of past inputs and observed states \citep{ljungDeepLearningSystem2020}.
While recurrent neural networks have mostly focused on modeling dynamics during task stimuli \citep{vyasComputationNeuralPopulation2020, manteContextdependentComputationRecurrent2013}, they may also  investigate how dynamic working-memory mechanisms can impact variations in time-dependent responses to sequences of targeted or sensory stimulation. 
For example, a comparison of recurrent neural networks with other memoryless neural network architectures demonstrated the necessity of internal memory mechanisms in predicting responses to visual stimuli sequences of varying images \citep{gucluModelingDynamicsHuman2017}.

In summary, in this section we have explored the variety of different generative models, as well as their different aims, advantages, and disadvantages for modeling dynamics evoked by targeted stimulation.
Choosing an appropriate model is crucial for investigating different types of questions about this form of dynamics: testing mechanisms related to the physiology of neurons would benefit from a biological neural network, or exploring the role of the connectome in the spatially distributed response to TMS could benefit from a phenomenological linear network model, or a project to accurately predict the firing rates of single neurons in response to optogenetic stimulation could benefit from system identification models or artificial neural network models.
Integrated with datasets of activity from targeted stimulation experiments, generative models complete our toolkit for uncovering the mechanisms underlying the complex, high-dimensional dynamics measured from this unique experimental paradigm.

\section{Conclusion}

Emerging targeted stimulation techniques such as optogenetics and electrode stimulation, combined with measurement techniques such as calcium imaging and EEG, now allow us to measure the intricate spatiotemporal responses of neurons and neural populations to a range of perturbations with unprecedented precision \citep{deisserothOptogenetics10Years2015, kellerCorticocorticalEvokedPotentials2014, rogaschAssessingCorticalNetwork2012, rothDREADDsNeuroscientists2016}.
Using an abstracted picture of brain dynamics, and drawing on results from dynamical systems theory, we presented an argument for why the dynamics from novel states accessed in these targeted stimulated experiments may be driven by new types of mechanisms, distinct to those underpinning the simpler low-dimensional dynamics of spontaneous activity.
The mechanisms which govern dynamics evoked by targeted stimulation are also functionally significant, as they reflect how the brain recruits diverse neural populations while processing complex streams of sensory inputs.
Fortunately, through an integration of targeted stimulation and measurement techniques with different generative models, we can uncover, analyze, and better understand these mechanisms at a chosen spatial scale.
Targeted stimulation techniques allow us to precisely perturb the brain to a range of states distant from the attractor manifold on which spontaneous dynamics unfold.
Measurement techniques can simultaneously capture the brain's trajectory as it relaxes back to the attractor, generating datasets of activity that could reveal the novel (and likely highly nonlinear) mechanisms that govern the relaxation's dynamics.
Finally, by integrating datasets of activity evoked by targeted stimulation with generative models, it is possible to rigorously test and validate mechanisms that we hypothesize to drive the dynamics.
Different generative modeling approaches are suited to testing different types of hypothesized mechanisms, from different physiological processes that can be incorporated in physiological models \citep{breakspearDynamicModelsLargescale2017, decoDynamicBrainSpiking2008}, to the abstracted dynamical mechanisms treated in phenomenological models \citep{cabralFunctionalConnectivityDynamically2017, kimLinearDynamicsControl2020}, and the statistical mechanisms empirically learned using data-driven models \citep{ljungSystemIdentification1998, ljungDeepLearningSystem2020}.
Our exploration touches on some of the many opportunities to uncover mechanistic insights into complex brain dynamics, ultimately guiding the practical implementation of emerging neuromodulation technologies in both research and clinical settings.

While our dynamical systems framework provides a foundation for interpreting the unique characteristics of brain dynamics evoked by targeted stimulation compared to spontaneous brain activity, there are alternative ways to conceptualize these dynamics. 
One such approach involves the separation of timescales \citep{kuehnIntroduction2015}. 
Some mechanisms may act too quickly to significantly influence the trajectories of fluctuations near the spontaneous activity attractor but become crucial in capturing the far-from-attractor dynamics evoked by targeted stimulation.
Brain activity may therefore possibly be modeled as a multi-timescale system, with mechanisms operating over various timescales in response to targeted stimulation \citep{babaie-janvierNeuralFieldTheory2020, chaudhuriLargeScaleCircuitMechanism2015}.
Targeted stimulation can also trigger new forms of non-equilibrium dynamics if multiple attractors exist in the state space, where specific perturbations can cause the relaxation trajectory to migrate to different attractors.
Each attractor may represent unique forms of spontaneous brain activity, such as the states of deep sleep versus wakefulness \citep{decoAwakeningPredictingExternal2019}, or sequential discrete thoughts during cognition \citep{spiveyContinuousDynamicsRealTime2006}.
Generative models of targeted stimulation can therefore inform the development of neuromodulation treatments to restore healthy brain function, by designing optimal perturbations that can force brain states to transition from `diseased' to `healthy' attractor dynamics \citep{muldoonStimulationBasedControlDynamic2016, perlPerturbationsDynamicalModels2021, kringelbachBrainStatesTransitions2020}.

As a novel experimental paradigm, datasets generated from targeted stimulation experiments allow us to not only test newly hypothesized mechanisms of brain dynamics, but also to refine existing mechanisms which have been validated mostly on datasets of spontaneous activity. 
Integrating existing models with new datasets can potentially contribute towards resolving standing issues of degeneracy in computational neuroscience, where generative models founded on differing mechanisms can lead to similar dynamics of spontaneous activity \citep{bernardBrainsBestKept2023}.
It is possible that the relaxation dynamics measured in targeted stimulation experiments, given their potential nonlinearities, can be used to arbitrate between different generative quantitative models of brain dynamics, beyond the relatively simple statistical properties of spontaneous activity that both forms of models are known to adequately capture, such as their pairwise linear correlation structure (as ‘functional connectivity’) \citep{nozariMacroscopicRestingstateBrain2023, messePredictingFunctionalConnectivity2015}.

Looking at and beyond the targeted stimulation setting, we suggest that more synergistic interactions between experimentalists and theoreticians can accelerate progress in our mechanistic understanding of brain dynamics.
As experimentalists forge more sophisticated paradigms that each generate unique forms of brain dynamics, theoreticians can build more robust generative models that can explain or predict the dynamics of brain activity from a wider range of experimental settings.
Conversely, refined generative models motivate the development of further experimental designs tailored to test and validate new theoretical accounts and predictions.
This reciprocal relationship between experimental precision and theoretical rigor can uncover novel hypotheses and validate theoretical predictions, fostering a deeper and more integrative comprehension of the intricate mechanisms underlying brain function.

\section{Acknowledgments}
The authors thank Nigel Rogasch for feedback and suggestions for the manuscript. 
RM would like to thank the Australian Government Research Training Program (RTP) Scholarship for financial support. 
BDF would like to thank the Selby Scientific Foundation for financial support.

\newpage

\bibliographystyle{unsrtnat_et_al}
\bibliography{bibliography}

\end{document}